\documentclass[aps,prl,twocolumn,superscriptaddress,groupedaddress]{revtex4}
\usepackage{graphicx}  
\usepackage{dcolumn}   
\usepackage{bm}        
\usepackage{amssymb}   
\usepackage{xcolor}	   
\usepackage{placeins}

\begin{document}



\title{Dominant $s$-wave superconducting gap in PdTe$_2$ observed by tunneling spectroscopy on side-junctions}
\author{J.A.~Voerman}\thanks{These two authors contributed equally to the work} \affiliation{MESA+ Institute for Nanotechnology, University of Twente, 7500 AE Enschede, The Netherlands}
\author{J.C.~de Boer}\thanks{These two authors contributed equally to the work} \affiliation{MESA+ Institute for Nanotechnology, University of Twente, 7500 AE Enschede, The Netherlands}
\author{T.~Hashimoto}\affiliation{MESA+ Institute for Nanotechnology, University of Twente, 7500 AE Enschede, The Netherlands} \affiliation{Yukawa Institute for Theoretical Physics, Kyoto University, Kyoto 606-8502, Japan}
\author{Yingkai~Huang} \affiliation{Van der Waals-Zeeman Institute, University of Amsterdam, Science Park 904, 1098 XH Amsterdam, The Netherlands}
\author{Chuan~Li}\affiliation{MESA+ Institute for Nanotechnology, University of Twente, 7500 AE Enschede, The Netherlands}
\author{A.~Brinkman}\affiliation{MESA+ Institute for Nanotechnology, University of Twente, 7500 AE Enschede, The Netherlands}
\vskip 0.25cm
\date{\today}

\begin{abstract}
We have fabricated superconductor-normal metal side-junctions with different barrier transparencies out of PdTe$_2$ crystalline flakes and measured the differential conductance spectra. Modeling our measurements using a modified Blonder Tinkham Klapwijk (BTK) formalism confirms that the superconductivity is mostly comprised of the conventional $s$-wave symmetry. We have found that for junctions with very low barrier transparencies, the junctions can enter a thermal regime, where the critical current becomes important. Adding this to the BTK-model allows us to accurately fit the experimental data, from which we conclude that the superconductivity in the $a$-$b$ plane of PdTe$_2$ is dominated by conventional $s$-wave pairing. 
\end{abstract}

\pacs{}
\maketitle

\section{I. Introduction}
The search for the elusive Majorana particle has brought physicists to the area of topological superconductivity. The mixture of Dirac physics and superconductivity (SC) is seen as a promising way of creating Majorana quasiparticles \cite{fu2008}, which in turn opens up the possibility of quantum computing through a process called braiding \cite{dassarma2015}. Experimental research has focused on the interface effects of superconductors coupled to either semiconductors with strong spin-orbit coupling \cite{mourik2012,rokhinson2012} or topological matter \cite{wiedenmann2016,sun2016,snelder2013,li2017}. Topological superconductors, for example Cu$_x$Bi$_2$Se$_3$, are also studied in the context of Majorana physics \cite{fu2010,sasaki2011}. The transition metal dichalcogenide PdTe$_2$ belongs to the P$\bar{3}$m1 space group and is known to be a superconductor \cite{leng2017,das2018,amit2018}. Recent experiments have shown that this material is also topological as it possesses a type-II Dirac cone \cite{liu2015,noh2017,clark2018}, highlighting it as an extraordinary material, that could host unconventional superconductivity intrinsically \cite{sato2017}. Notably, Teknowijoyo \textit{et al}. have narrowed the possible order parameter (OP) symmetries down to three candidates: A$_{1g}$ (conventional $s$-wave) pairing, A$_{1u}$ (helical $p$-wave) pairing, or E$_{u(1,0)}$ (nematic $p$+$f$-wave) pairing, by showing that the order parameter of PdTe$_2$ is nodeless \cite{teknowijoyo2018}. The latter two pairings are nontrivial. Experiments investigating the nature of the superconductivity in PdTe$_2$ have so far found no indication of unconventional superconductivity \cite{das2018}.

\begin{figure*}
\centering
\includegraphics[width=.9\textwidth]{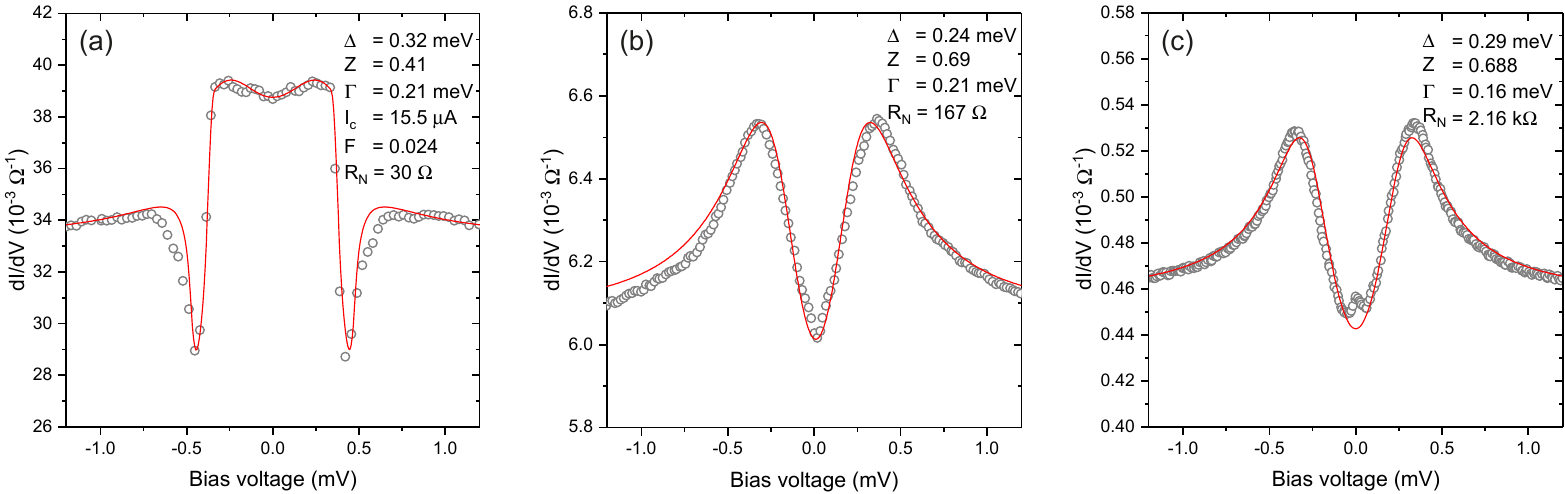}
\caption{\label{fig:basetemp} $dI/dV$ spectra of three PdTe$_2$ junctions with different resistances measured at base temperature (grey circles). The red line is our best fit to the data. All relevant fitting parameters, as well as $R_N$, are included in the panel. (a) $dI/dV$ measurements and fit of a BTK-model with very transparent PdTe$_2$/Au interface ($R_N$ = 30 $\Omega$) and critical current effects. (b) $dI/dV$ measurements and fit of a BTK-model with slightly less transparent PdTe$_2$/Al$_2$O$_3$/Pd interface ($R_N$ = 167 $\Omega$). (c) $dI/dV$ measurements and fit of a BTK-model with an opaque PdTe$_2$/Al$_2$O$_3$/Au interface ($R_n$ = 2.16 k$\Omega$).}
\end{figure*}

In this article we present tunneling spectroscopy measurements performed on PdTe$_2$-normal metal side-junctions, to shed light on the in-plane properties of the order parameter and distinguish between the three possible OPs that Teknowijoyo \textit{et al}. have singled out. We model the data using a combination of the Blonder Tinkham Klapwijk (BTK) formalism \cite{btk} and the effect of the critical current, $I_c$, on the differential conductance \cite{sheet2004,baltz2009,daghero2010,naidyuk2018,taboryski1996,verkin1979}. Finally we show additional features found in the $dI/dV$ spectrum of the purely ballistic junction, together with an analysis of these features that are beyond our BTK and $I_c$-model. 

\section{II. Experimental Details}
We have fabricated our superconductor-(insulator-)normal metal [S(I)N] junctions out of exfoliated flakes of a PdTe$_2$ crystal. The crystal has a preferred cleavage plane, which orientates all flakes with the $c$-axis out of plane. The single crystal of PdTe$_2$ was grown by a modified Bridgman method. High purity Pd (99.99\%) and Te (99.9999\%) were used as starting materials. The desired components were sealed in an evacuated cone-ended quartz ampoule. The ampoule was heated up to 800 $^\circ$C, kept for 48 hours and then cooled down to 500 $^\circ$C at a rate of 3 $^\circ$C per hour, followed by furnace cooling.

All devices are prepared by Ar$^+$ milling through the flake prior to the deposition of a barrier and normal metal, in order to create a side-contact, allowing us to probe the in-plane properties of the superconducting order parameter. All patterning for these steps was done using standard electron-beam lithography. The devices differ in their interfaces between the PdTe$_2$ and the normal metal. The first type of devices was made without a specific barrier and is a 500 nm wide SN interface between PdTe$_2$ and gold, with a normal state resistance ($R_N$) of about 30 $\Omega$ at 15 mK. The second type of devices was made by transferring the argon milled flakes to a sputter machine where they were cleaned of contaminations by low RF power plasma etching. On the cleaned surface, 1 nm of Al was sputter deposited, followed by oxidiation in 10 mbar of oxygen for one hour to form an Al$_2$O$_3$ oxide barrier. To finalize the devices, a normal metal layer of palladium was sputter deposited on the aluminium oxide without breaking the vacuum. The $R_N$ was about 200 $\Omega$ at 15 mK. Of the third type of devices only one was fabricated. This device was transferred to an atomic layer deposition (ALD) apparatus after argon milling, where a 1.2 nm thick Al$_2$O$_3$ layer was grown at 100 $^{\circ}$C, followed by ex-situ deposition of 40 nm of gold by sputter deposition. These SIN junctions have an $R_N$ of about 2 k$\Omega$ at 15 mK. Although many junctions were made, one SIN-junction had this resistance value at base temperature, whereas the other showed R $>$ 1 M$\Omega$. Several of our similarly fabricated SIS-junctions did show an R$_N$ of 2 k$\Omega$ at 15 mK, but are not included in this work on SIN junctions. For each of the three S(I)N types, measurements on one representative device are presented in this work.\\

\section{III. Results and analysis}

We drive a DC bias current with a small AC excitation through the junctions while measuring both the DC and AC response across the junction to probe the differential resistance.
The measured differential resistance is numerically inverted to differential conductance and plotted against the measured DC bias voltage. The results of these measurements at the lowest temperature reached ($T<$100 mK) are shown as grey circles in figure \ref{fig:basetemp}. Comparing the three graphs we see a clear evolution of the main feature around zero bias.
In figure \ref{fig:basetemp}(a) we see a dented plateau around zero bias, accompanied by sharp dips in conductance at $\pm$0.5 mV. Figure \ref{fig:basetemp}(b) shows a quite different shape. The dented plateau around zero has been replaced by an Andreev like spectrum with coherence peaks surrounding clear dip. 
The final device, whose differential conductance is shown in figure \ref{fig:basetemp}(c), has the highest normal state resistance. 
Just as the data in figure \ref{fig:basetemp}(b), the measured $dI/dV$ spectrum in \ref{fig:basetemp}(c) looks like a clear Andreev spectrum. Around zero bias a small zero bias conductance peak (ZBCP) is visible.\\

\begin{figure*}
\centering
\includegraphics[width=.8\textwidth]{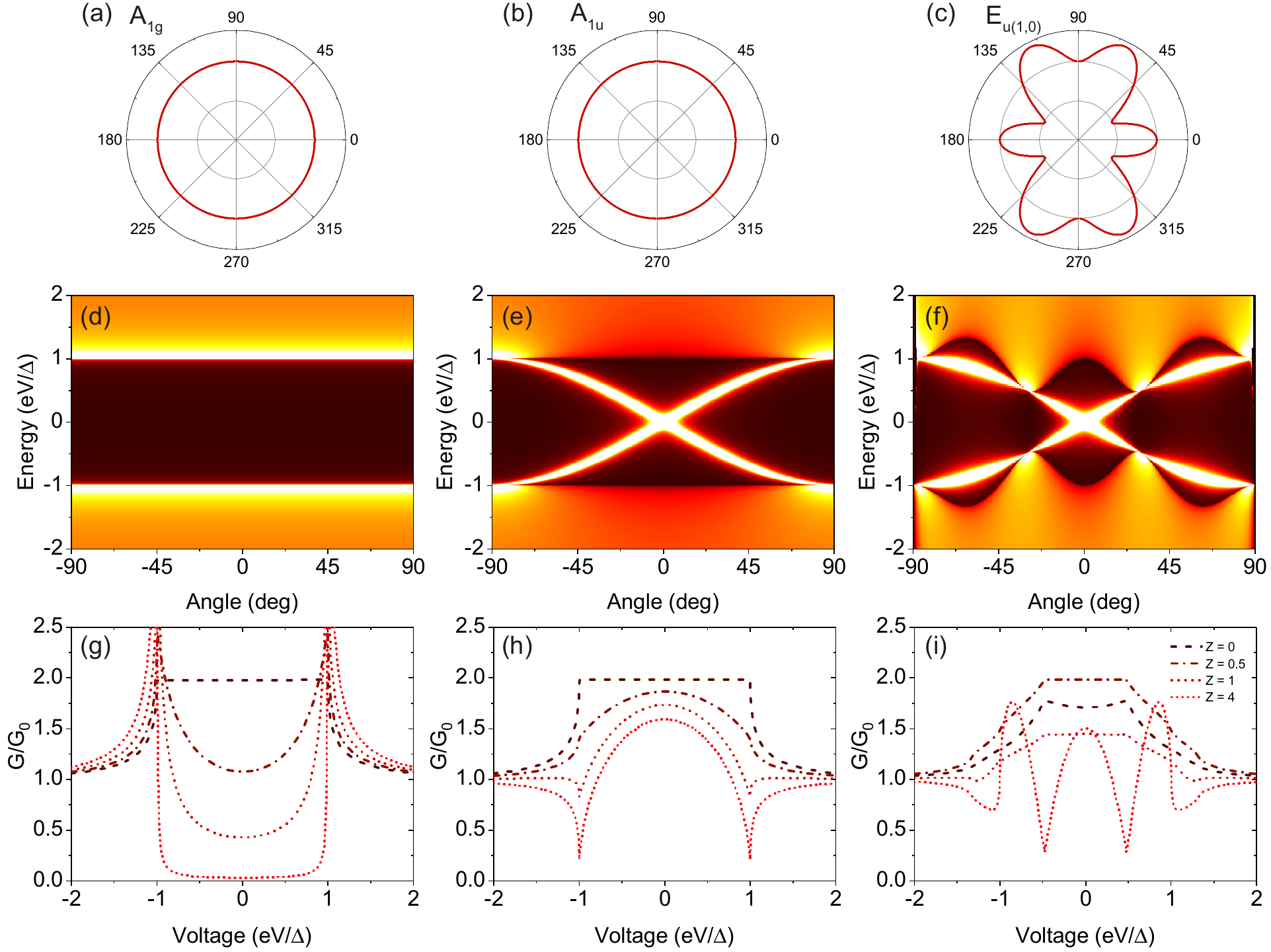}
\caption{\label{fig:OPModel}The BTK-model for conventional $s$-wave symmetry, A$_{1u}$, and E$_{u(1,0)}$ pairing. (a)-(c) show the angle dependence of the gap, $\Delta$, indicating the shape of the OP and the fact that it is nodeless. (d)-(f) show the angle dependent conductance at different energies calculated for barrier strength $Z$ = 4. The colorscale reflects the conductance, where brighter colors indicate higher conductance. Both A$_{1u}$ (e), and E$_{u(1,0)}$, (f) have helical edge states at zero energy. Note that due to the anisotropy of the E$_{u(1,0)}$ pair potential, some of the states with large $k_y$ components on the normal metal side have no superconducting equivalent at the same energy and result in zero conductivity. (g)-(i) are the conductance spectra obtained for different dimensionless barrier strengths, $Z$, in the BTK-model. They are the result of averaging the conductance over angles between -90 and +90 degrees. The legend in (i) shows which line represents which $Z$ and is valid for (g) and (h) as well. }
\end{figure*}

Because our PdTe$_2$ flakes are less than 100 nm thick, which is less than the reported superconducting coherence length \cite{leng2017,teknowijoyo2018}, we have modeled the conductance spectra numerically using a 2D BTK formalism for different order parameters. In this model, the bands are assumed to be parabolic since the Fermi energy is much larger than the energy where the type-II Dirac points reside \cite{liu2015}. The chemical potential mismatch, $\mu_{sc}/\mu_n$, is set to 1 for simplicity, so that $Z = H m_e /\hbar^2 k_{sc}$, with $H$ the height of the $\delta$-shaped barrier, is the only barrier parameter. Teknowijoyo \textit{et al}. have experimentally determined the OP in PdTe$_2$ to be nodeless, which, together with crystal symmetry constraints, leaves us with three different pair potentials: A$_{1g}$ (conventional $s$-wave), A$_{1u}$ (helical $p$-wave), and E$_{u(1,0)}$ (nematic $p$+$f$-wave). The latter two correspond to the d-vectors $\mathbf{d}_{A_{1u}} = k_x\hat{x}+k_y\hat{y}+k_z\hat{z}$ and $\mathbf{d}_{E_{u(1,0)}} = k_x(k_x^2-3k_y^2)\hat{x}+k_z\hat{y}+k_y\hat{z}$. Before moving on to fitting our measured conductance spectra, we show the general features of the three OPs in a 2D BTK-model, where $k_z = 0$. Figures \ref{fig:OPModel}(a)-(c) show the angle dependence of the superconducting gap magnitude. It is obvious from these plots that all three are nodeless. Note that OPs of the $s$-wave, (a), and A$_{1u}$ pairing, (b), differ in their angle-dependence of the phase, rather than gap magnitude. Figures \ref{fig:OPModel}(d)-(f) show the normalized conductance as a function of the angle with respect to the interface normal. Brighter colors indicate a higher conductance in these graphs. Both the A$_{1u}$ and the E$_{u(1,0)}$ pairing exhibit helical edge states within the superconducting gap. The panels labelled (g)-(i), show the calculated conductance spectra for dimensionless barrier strength $Z$ = 0, 0.5, 1, and 4. The legend is included in figure (i). These $dI/dV$ spectra are the result of angle averaging over a semicircle directed at the interface. 
It should be noted that no signs of unconventional superconductivity have been found in differential conductance measurements along the $c$-axis, which rules out 3D isotropic A$_{1u}$ pairing but leaves room for an anisotropic variant \cite{das2018,clark2018}. The other nodeless pairing symmetry, E$_{u(1,0)}$, consists of components that are linear in $k$ and cubic in $k$, i.e. $p$-wave + $f$-wave symmetry. This can behave like a fully gapped system only when the $k^3$ component is sufficiently strong compared to the linear term. Although the E$_{u(1,0)}$ nematic $p$+$f$-wave state is unlikely to occur in nature, recent reports on the topological superconductor Cu$_x$Bi$_2$Se$_3$ have found indications of E$_u$ pairing symmetry \cite{matano2016,hashimoto2013,yonezawa2017}.\\ 

Comparing our model to the differential conductance curves displayed in figure \ref{fig:basetemp}, it appears that only conventional $s$-wave pairing can not adequately explain all our findings. Although the data obtained on the two high resistance devices 
can be nicely replicated using $s$-wave pairing with some degree of broadening, the sharp dips and elevated plateau of the lower resistance device are absent in figure \ref{fig:OPModel}(g), which shows the resulting differential conductance of the $s$-wave BTK-model. The helical $p$-wave $dI/dV$, figure \ref{fig:OPModel}(h), does exhibit sharp dips and a rising plateau, albeit far more rounded than the experimental data. Unconventional superconductivity, as long as it is nodeless \cite{teknowijoyo2018}, is not unimaginable in PdTe$_2$ as the unique spin (or pseudo spin) structure of a Dirac semimetal (DSM) can stabilize unconventional pairing mechanisms \cite{fu2010,sato2017}.
However, using a combination of conventional $s$-wave and helical $p$-wave pairing, we were unable to accurately model the data of figures \ref{fig:basetemp}(a), i.e. the most transparent junction. The differential conductance of the two most resistive junctions, on the other hand, can be fitted well using only the conventional $s$-wave pairing model. The result of this fitting procedure is shown as a red line in figures \ref{fig:basetemp}(b) and (c). Every part, except for the ZBCP in \ref{fig:basetemp}(c) is described by the BTK-model, using a dimensionless barrier strength $Z \approx$ 0.7. We will leave the ZBCP for now and turn our attention to the lower $R_N$ junctions.\\

\begin{figure}
\centering
\includegraphics[width=.48\textwidth]{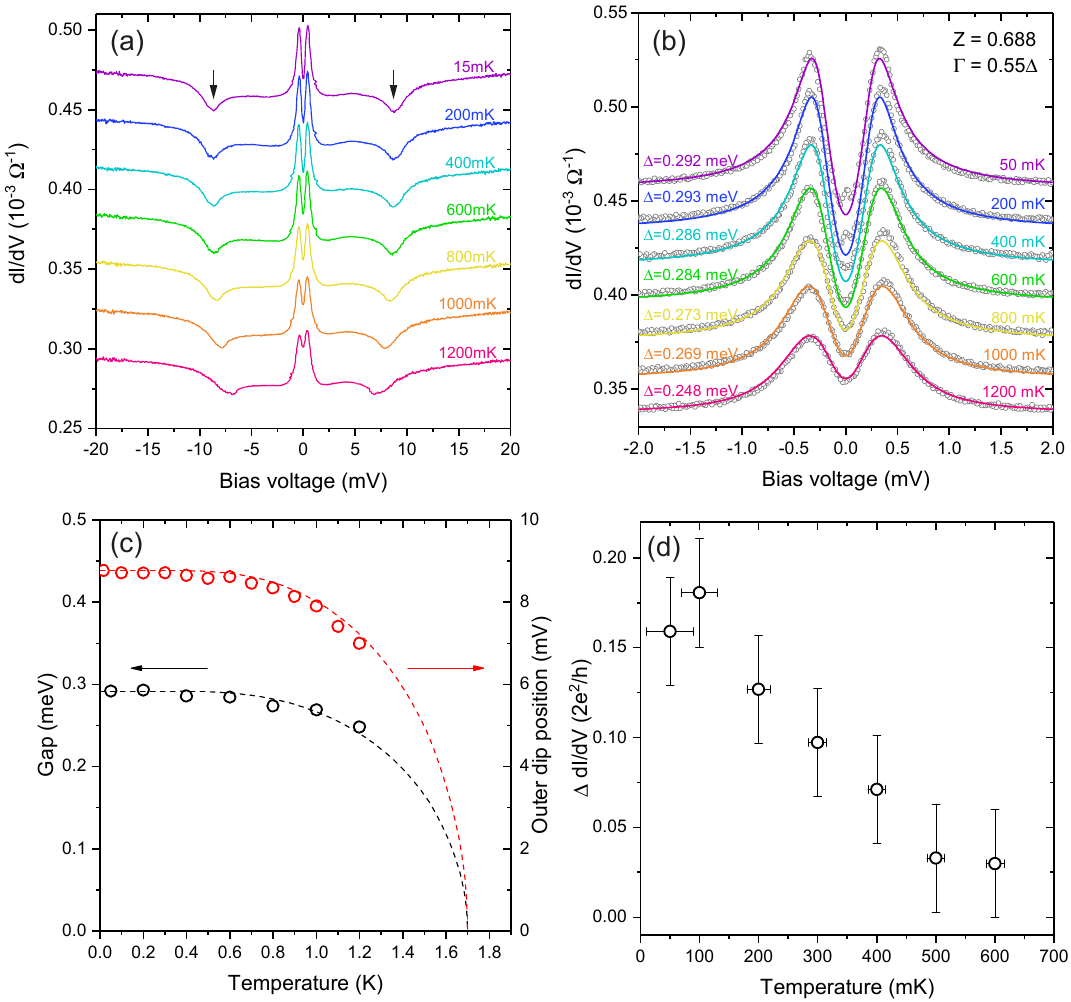}
\caption{\label{fig:ptopbonus} Additional measurements and analysis on the highest resistance sample. (a) $dI/dV$ of the 2.16 k$\Omega$ sample for different temperatures measured over a large range of bias voltage. For clarity, all the curves except for the 15 mK curve have been given a constant offset. (b) $s$-wave BTK fits (colored lines) to the measured $dI/dV$ of the 2.16 k$\Omega$ junction (grey circles) at different temperatures. Again, all the curves except for the 15 mK curve have been given a constant offset. The temperature and fitted gap $\Delta$ are indicated next to the line. $Z$ and $\Gamma$ are shared across the curves and are indicated in the top-right corner of the graph. (c) The superconducting gap from the BTK fits as a function of temperature (black circles) and the position of the shallow dip versus temperature (red circles). Dashed lines show the standard temperature dependence from BCS theory for a $T_c$ of 1.7 K. (d) The height of the ZBCP as a function of temperature.}
\end{figure}

To explain the origin of the sharp dips and dented plateau of figure \ref{fig:basetemp}(a) we extend our BTK-model by taking into account the influence of the interfacial critical current on the obtained $dI/dV$ spectra \cite{sheet2004,baltz2009,daghero2010,naidyuk2018,taboryski1996,verkin1979}. See Supplemental Material for additional information on the model. The BTK-model assumes ballistic transport through the junction at all bias voltages, but in the case of very low resistance junctions, the junction may leave the ballistic regime and enter the thermal regime: while increasing the bias current through these transparent junctions we reach the critical current $I_c$, an effect that comes on top of the BTK-model. This critical current does not refer to the typical bulk $I_c$ of a superconductor, but rather to a reduced critical current in the disordered surface of the PdTe$_2$ close to the interface. Upon reaching this critical current, a voltage suddenly appears across the junction, which is represented as a step in the IV-characteristic. Taking the derivative of this will yield sharp dips in $dI/dV$ at the critical current. The red line in figures \ref{fig:basetemp}(a) shows the striking agreement of an $s$-wave BTK-model with our data, when the effect of $I_c$ is taken into account. The BTK parameters, as well as $I_c$, are reported inside the graph. $F$ describes the mixing of the $I_c$-model and the BTK-model through a linear combination, where $F$ = 0 means purely BTK-model and $F$ = 1 is purely the $I_c$-model. We stress that this fitting is performed using only conventional $s$-wave pairing, like in figure \ref{fig:basetemp}(c). The device with the largest barrier is apparently in the ballistic regime for all applied currents, whereas this does not hold for the most transparent device. The $dI/dV$ features, occasionally ascribed to unconventional superconductivity, arise in our case from high transparency of the junction in combination with a disordered interface. This high transparency can be due to the design of the device, or an accidental feature, such as a pinhole or otherwise broken barrier. For an example of how such a $dI/dV$ spectrum can easily be misinterpreted as originating from $p$-wave superconductivity see Supplemental Material.\\

So far we have found that the main features of all three measured devices can be understood with the same OP symmetry, even though their differential conductance spectra differ greatly. The high resistance device is our best candidate for a more extensive conductance spectroscopy study, as it exhibits ballistic transport for up to 8 mV. We have measured the $dI/dV$ spectrum of this device at different temperatures and for a much larger range of bias currents. The experimental data is presented in figure \ref{fig:ptopbonus}(a). Figure \ref{fig:ptopbonus}(b) serves as a zoomed in version of this graph around zero bias and shows, as solid lines, the curves obtained from a conventional $s$-wave BTK-model. The theoretically obtained curves describe the data very well. A shallow dip-feature presents itself in figure \ref{fig:ptopbonus}(a) at bias voltages greater than 5 mV, far beyond the superconducting gap ($\sim$300 $\mu$V). The dip differs from the dips earlier attributed to the critical current. Such dips are quite sharp, since they relate to an instant increase in voltage, whereas this feature is shallow and stretched wide in voltage. Furthermore, we have plotted the position of this dip as a function of temperature in red circles in figure \ref{fig:ptopbonus}(c). They are accompanied by the superconducting gap $\Delta$ as extracted from the BTK fit on the low-bias part of this dataset. Both temperature dependences can be described using standard BCS theory. The two dashed lines show this standard BCS behavior, scaled to the voltage value at the lowest temperature. 

Over the past decades there have been numerous experiments in which dips such as presented in figure \ref{fig:ptopbonus}(a) have been observed \cite{nguyen1992,raychaudhuri2004,bashkalov2005}. A possible origin of this dip is that weak spots in the barrier are responsible for the crossover into the thermal regime at larger currents, similar to our I$_c$-model \cite{raychaudhuri2004,bashkalov2005}.\\

The final feature we discuss is the aforementioned ZBCP. This peak can clearly be distinguished in low bias region of the low transparency device at sufficiently low temperatures. We have subtracted the BTK fits shown in figure \ref{fig:ptopbonus}(b) from the respective data and tracked the height of this peak as a function of temperature. The extracted peak heights are shown in figure \ref{fig:ptopbonus}(d). We see that the temperature dependence of the ZBCP is linear and the temperature at which the peak appears is much lower than the reported critical temperature of PdTe$_2$. At these temperatures the thermal energy, $k_B T$, is smaller than the width of the ZBCP, which hints that we are probing a different characteristic energy scale here. Recent studies of the superconductivity of PdTe$_2$ have confirmed the existence of multiple superconducting channels, related to parallel bulk and surface superconductivity \cite{leng2017}. Because only one device of the third type was fabricated, it is unclear whether the ZBCP arises from intrinsic superconducting properties, or from interface effects inside this specific device. \\ 

\section{IV. Conclusions}

In short, we have fabricated three PdTe$_2$/Normal metal side-junctions with different transparencies. The shape of the conductance spectra heavily depends on the normal state resistance of the junction and can make conventional superconductivity look unconventional. One should exert caution in analyzing the data of low resistance SN junctions and confirm that the junction is solely in the ballistic limit, or include the effect of the critical current on the $dI/dV$ spectrum in one's model. Taking these critical current effects into account in the data analysis, the conductance spectroscopy measurements on our devices indicate that the order parameter in PdTe$_2$ is dominated by conventional $s$-wave pairing.

\section{\label{sec:ack}Acknowledgments}
This work was financially supported by the European Research Council (ERC) through a Consolidator Grant. T.H. is supported by the JSPS KAKENHI (No. JP15H05855).

\setcounter{figure}{0}
\renewcommand{\theequation}{S\arabic{equation}}
\renewcommand{\thefigure}{S\arabic{figure}}
\renewcommand{\bibnumfmt}[1]{[S#1]}
\renewcommand{\citenumfont}[1]{S#1}
\renewcommand{\thetable}{S\arabic{table}}

%
%

\clearpage

\onecolumngrid
\begin{center}
  \textbf{\large Supplementary information to: Dominant $s$-wave superconducting gap in PdTe$_2$ observed by tunneling spectroscopy on side-junctions}\\[.2cm]

\end{center}
\twocolumngrid

\section{A. A minimal model for low resistivity N(I)S devices}
\begin{figure}
\centering
\includegraphics[width=8cm]{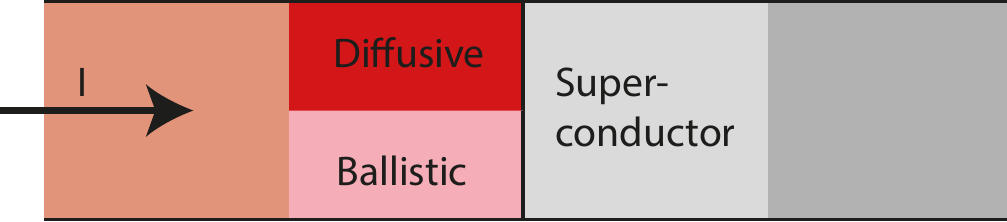} 
\caption{\label{fig:schematic} A schematical illustration of the interface in this model. The left side represents the metal side, which is described by parallel ballistic and diffusive channels, and the right represents the superconductor side of the interface.} 
\end{figure}

When current-biasing low resistivity devices, one often needs to send relatively high currents to reach voltages of the order of the gap magnitude. In the case of a disordered N-S interface, the critical current of the superconductor near the interface can be reached before $e V_{bias}\approx\Delta$ and the resulting $dI/dV$ spectrum is no longer well described by the ballistic BTK model. In figure \ref{fig:schematic}, a typical low resistance N(I)S interface is described as a metal side, which has a diffusive and a ballistic channel, in series with a superonducting side, which we assume to be ballistic in the normal state for this minimal model. The ballistic and diffusive contributions of the metal are modeled as $W$ and $1-W$ respectively, with $W$ a fitting parameter between 0 and 1.

The $I$-$V$ characteristic of a diffusive metal can in principle be easily described as a linear dependence $V_{dm} = I R_{m}$. This diffusive contribution causes heating at the interface as $T_{eff}^2 = T_{bath}^2 + V_{bias}^2/4L$, with $L$ the Lorentz number \cite{verkin1979}. Because of the small temperature dependence $R(T)$ of metals at low temperatures, we assume the metal resistance to be constant. The ballistic $I$-$V$ characteristic of the metal is also taken into account as $V_{bm} = I R_{m}$, but without the aforementioned heating effect. The $I$-$V$-characteristic of a superconductor can generally be described as $V_{SC} = sgn(I) \, \Re(\sqrt{I^2-I_c^2}) R_{SC}$, which holds below and above $I_c$.

\begin{figure*}
\centering
\includegraphics[width=12cm]{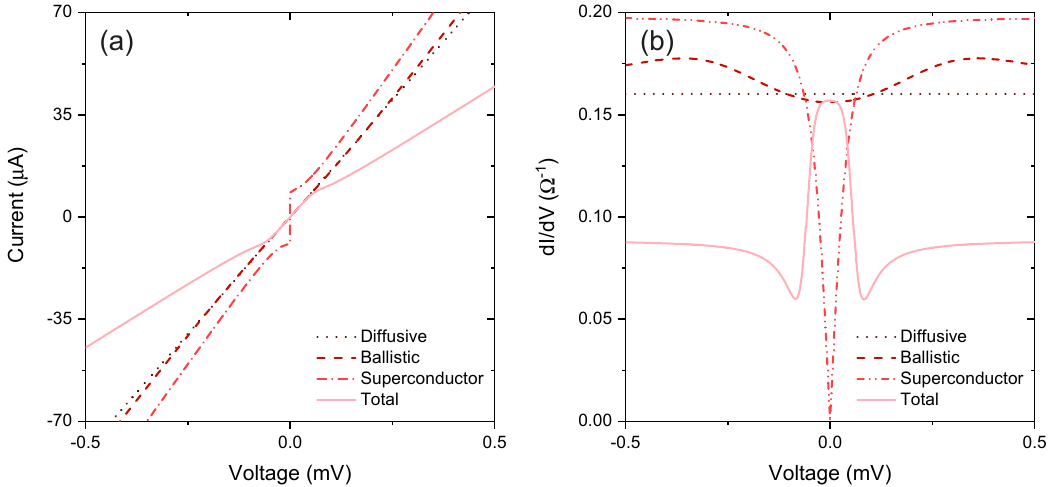} 
\caption{\label{fig:IVpanels} (a) $I$-$V$ characteristics of the separate parts of the interface, plotted against the voltage over this separate part. The total is modeled assuming a current bias setup. (b) The $dI/dV$ conductance spectra corresponding to the $I$-$V$ characteristics in panel a.}
\end{figure*}

For $I<I_c$, when the resistance contribution of the superconductor drops to zero, the resistance of the interface is entirely governed by the diffusive metal, which behaves again as $V_{dm} = I R_{m}$, and the ballistic metal which in this case is described by the BTK model: $V_{bm} = I R_{BTK}$ \cite{btk}. The resulting $I$-$V$ characteristics are shown in Figure \ref{fig:IVpanels}(a).

The resulting total conductance spectrum will follow BTK behavior as long as $I<I_c$, damped by a parallel, diffusive contribution, and will jump at $I=I_c$ to saturate at a constant value that corresponds to the total metal and normal state PdTe$_2$ resistances. The effective temperature $T_{eff}$ at the interface causes thermal smearing of the total $I$-$V$ characteristic. In the BTK picture, this thermal smearing is taken into account as a broadening of the bias voltage. For the critical current features however, thermal smearing should be taken into account in the $V$-$I$ curve by broadening the bias current $I_{Thermal} \sim V_{Thermal}/R_{total}$. This makes the amount of rounding of the critical current strongly dependent on the total resistance of the device. Figure \ref{fig:IVpanels}(b) shows the total $dI/dV$ conductance spectra, together with the conductance spectra of the isolated parts of the interface.

\section{B. An example of a low resistance device that mimics p-wave behavior}
\begin{figure*}
\centering
\includegraphics[width=\textwidth]{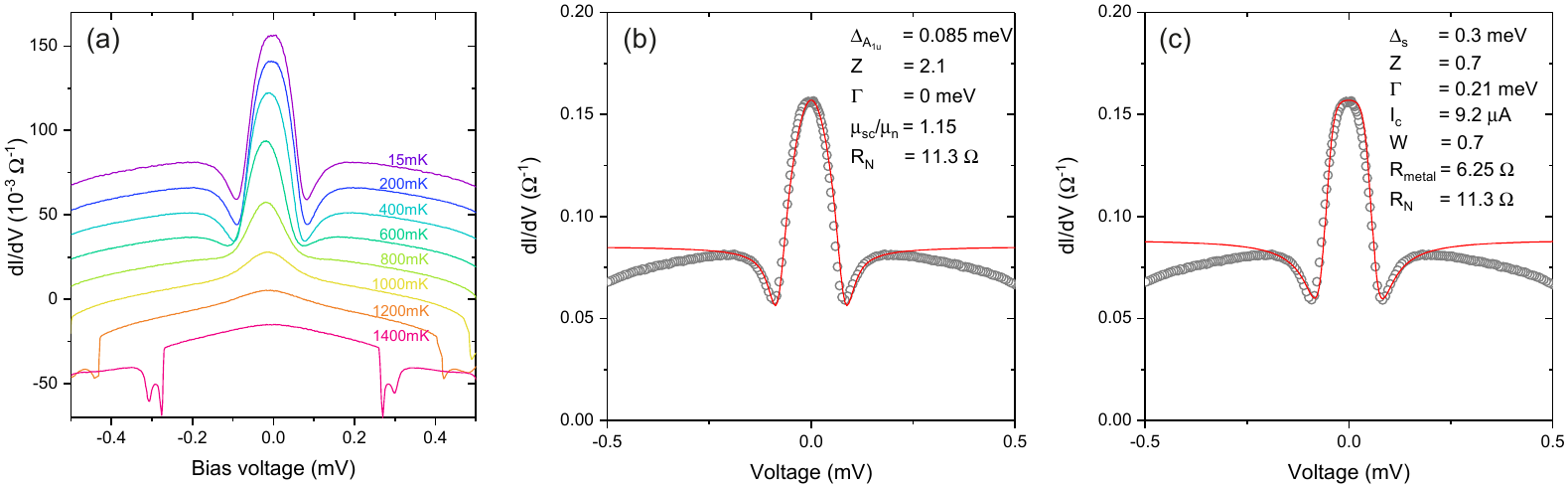} 
\caption{\label{fig:fits} (a) Temperature dependent $dI/dV$ spectra for an $R_n$ = 11.3 $\Omega$ N(I)S device. The curves, except for 15 mK, are offset for clarity. (b) A 2D, A$_{1u}$ BTK model fitted to the 15 mK data. (c) The $s$-wave + $I_c$ model, discussed in section A, fitted to the 15 mK data.}
\end{figure*}

One of the lowest resistance N(I)S devices fabricated for this work, has a resistance of 11.3 $\Omega$. The measured conductance specrum is shown in Figure \ref{fig:fits} and exhibits a pronounced dome at zero bias, along with a parabolic background that is associated with Joule heating effects \cite{baltz2009}. This spectrum perfectly matches what one would expect for a helical-$p$ superconductor with a high barrier (see Figure 2(h) of the main text). 

Figure \ref{fig:fits}(b) shows the excellent correspondence between the measured data and a numerical model for an A$_{1u}$ order parameter with $\Delta_{A_{1u}}$ = 0.085 meV, a dimensionless barrier strength Z = 2.1, and a small chemical potential mismatch $\mu_{sc}/\mu_n$ = 1.15. Despite the good fit to the data, the fitting parameters do not seem very appropriate for the device. Following the BCS model, the gap magnitude indicates $T_c \approx $560 mK, which does not match well to the temperature dependence in panel \ref{fig:fits}(a). The dimensionles barrier strength of Z = 2.1 is very high for a device with a normal state resistance of 11.3 $\Omega$. Contrary to the other datasets presented in the main text, interpreted as p-wave this spectrum shows no sign of an additional s-wave order parameter, which is another indication that these measurements should not be interpreted as fully described by the BTK formalism.

In figure \ref{fig:fits}(c), the critical current model described in section A is fitted to the data. The fit parameters indicate 70\% ballistic transport in the metal side of the barrier and a critical current $I_c$ = 9.2 $\mu$A. Instead of fitting the BTK parameters to the data, we assumed the same values as found from fits in the main text because the influence of the s-wave BTK contribution is rather small. This $s$-wave + $I_c$ model matches just as well to the data as the p-wave model does, but in this case with much more realistic fit parameters.

The experimental data and theoretical fits presented in this section illustrate neatly the difficulties of studying the superconducting order parameter with a low resistance, point-contact like setup. To make hard statements about the nature of the superconducting order parameter, one preferably uses devices with thick barriers without pinholes, so that the resulting tunneling spectrum accurately corresponds to the density of states of the superconductor.

\end{document}